# On the free, safe, and timely execution of component-based systems

Julio Cano and Marisol García-Valls

Dpto. de Ingeniería de Telemática/Universidad Carlos III de Madrid, Leganés, Spain
{jcano, mvalls}@it.uc3m.es

**Abstract.** Traditional real-time systems are reluctant to integrate dynamic behavior since it challenges predictability and timeliness. Current efforts are starting to address the inclusion of a controllable level of dynamicity in real-time systems to increase the degree of freedom or flexibility in the execution of real-time systems. This is mainly achieved by imposing a set of bounds and limitations to the allowed system structure and operations during a transition. The ultimate goal is to time-bound the duration and result of the system state transitions. One of the main obstacles for run-time transitions is the difficulty in characterizing the different operations and phases of a run-time system transition to guarantee a time bound for each phase. In this paper, an infrastructure is described to ensure the timely execution of state transitions that can be safely incorporated and performed at run-time in a component-based real-time system, preserving its temporal properties. The infrastructure allows to perform different management operations to modify components at run-time ensuring the overall schedulability of the system. The infrastructure is validated by its implementation in a specific component-based framework.

## 1 INTRODUCTION

Methods and techniques for the development of real-time systems traditionally had the goal of identifying the sources of uncertainty with the aim of constructing systems with time-predictable execution targeted at the safety critical real-time domain. Over the years, systems have evolved with the market demands and with the increased computational power of the new hardware. Such evolution has also reached the real-time domain in a way that techniques have been adapted to fit to the general purpose computing environments where timely operation is a clear added value. As a consequence, real-time techniques are progressively merged with other paradigms (e.g., distribution) that introduce a number of challenges such as uncertainty and dynamic behavior. Providing real-time execution guarantees in the presence of dynamic behavior requires the enforcement of some bounds and limitations in order to achieve feasible solutions.

Modern distributed systems benefit from a certain degree of dynamism. Currently, they are mostly built with service oriented paradigms and/or component technology providing loosely coupled software architectures where the entities (i.e., components or services) encapsulate the functionality pieces that can be, if needed, replaced by other entities or reconfigured. Whereas this is a conceptual point in favor, in practice, it becomes difficult to implement especially if the reconfiguration must also be seamless in the temporal domain, i.e., time-bounded in order not to cause any deadline miss.

Component technology is widely used to develop software-intensive embedded systems. The inclusion of real-time properties in component technology has been tackled by different approaches and even some proposals have become standards such as the OMG CORBA Component Model (CCM) [11] and later the MARTE profile [12] for real-time systems. Most of the real-time systems implementations, and also existing real-time component frameworks lack support for enabling *controlled dynamism* (e.g., see [3]). These component frameworks do not include some required features or operations for this like substitution of a component, deactivation of components that are not going to be used any longer in the system, or unloading the corresponding code to free resources. Further efforts must be made in current state of research to support adaptation of real-time and embedded component-based systems, so that they can support run-time changes to enable reconfiguration of components or the whole platform.

In this work, we refer to controlled dynamism as a set run-time changes [17] that are allowed in a system for enabling the change of some functionality due to a user request, a programmed event, or the need to recover from any detected failure. Resources need be provisioned by system for allocating such changes. We do not consider non-expected asynchronous changes in the system made by components themselves without any control nor provision from the system. We describe an infrastructure to support run-time execution of dynamic management operations in a real-time context. *Management operations* refer to the set of activities that can be performed in a dynamic component-based system, i.e., addition, removal, modification, or replacement of components. It relies on the existence of special tasks that undertake the execution of the management operations to enable the state transition by modification of the actual running component set. Relying on a previous work on a framework for scheduling of real-time reconfigurations [8][5], this framework includes an admission test for ensuring the feasibility of the specific management operations to be included.

The paper is structured as follow. Section 2 presents the related work on dynamic and real-time component-based systems. Section 3 describes the proposed infrastructure. Section 4 presents the run-time admission control and checks to determine the schedulability of management operations. Section 5 presents the validation of the infrastructure by means of its high-level specification and results of a prototype implementation including a summary discussion. Finally, section 6 concludes this paper by summarizing the achievements of the paper and describing the future work directions.

## 2   Related work

Different approaches have been used in the component-based software engineering literature in order to introduce real-time requirements and properties in component models. One of the most well-known component models is CORBA Component Model (CCM). Dynamic reconfiguration procedures are provided for CCM, like in [21] and [1]. The former looks for quiescence to be able to accomplish reconfigurations whereas the latter only blocks the directly affected components in the process.

Sharma et al. [20] also propose the use of components, called *qosket* components, to encapsulate the behavior of their Quality Objects (QuO) framework over the CCM, and they implement reusable adaptive behaviors. *Qoskets* are used as middleware components to encapsulate other components and to provide reconfiguration capabilities to distributed dynamic systems. However, no real-time capabilities are provided by this middleware framework. Another middleware framework that deals with distributed dynamic systems is described in [7]. This framework provides real-time capabilities to the distributed system.

Rasche and Polze, in [15] and [16], present a technique for dynamic reconfiguration of component-based software, in which components are blocked to apply management tasks such as reconfiguration of components. The reconfiguration is managed and applied in a way that the execution of all the current transactions between components is correct, but the application is blocked until the reconfiguration is finished without taking into account real-time deadlines.

Other approaches are in fact concerned about real-time properties as the one proposed by Wahler et al. in [22]. This model is mainly focused in a model for component replacements. This model may require several iterations to complete the replacement and does not that the copy of the component state is completed in a fixed number of periods given that the component is still working and its state is modified. This method also obviates the need to analyze the available slack time to make these component updates.

Real-Time Specification for Java (RTSJ) [2] is used to build real-time component-based frameworks and applications. However, dynamic adaptation issues are only treated by Plšek et al. in [14], where system adaptation and real-time component replacement issues are recognized as problems to be solved. No solution is currently provided for them.

Schneider et al. [19] introduce a proposal to make use of their real-time middleware OSA+ to be able to apply component replacements. Blocking, partially blocking, and non-blocking modes are proposed depending on how many components are blocked during the replacement process. Timing control of the replacement is supposed, but no exact temporal model is described for the replacement. Neither CPU time reservation nor interference of a replacement in the schedulability of the system is considered in this work.

Other work to provide QoS assurance during the reconfiguration is tackled by, for instance, [8] which proposed methods to reserve resources to enable system reconfiguration according to varying needs [5]. This method is oriented to multimedia systems and based on a budget scheduling model and dynamic priority assignment [6], where resources are reserved for every component independently so that these components can perform a reconfiguration to be adapted to the QoS requirements of the system. This method focuses on providing QoS assurance for a continuously reconfiguring system in multimedia domains.

There are also recent contributions on providing strictly correct configurations based on on-line verification methods such as [24], [25] and [26] in the domain of Cyber-Physical Systems. These can be embedded in the logic of a communication middleware for fast verification as proposed in [27].

iLAND (mIddLewAre for deterministic dynamically reconfigurable NetworkeD embedded systems)[7] is an ongoing research project whose objective is to develop a component-based middleware with deterministic dynamic functional composition and reconfiguration. Operating systems, service-oriented and real-time approaches are combined to achieve this objective. Real-time composition and reconfiguration is provided. But this reconfiguration is deterministically designed.

Whereas works like iLAND find their dynamism limited at design time, the proposed mechanism allows increasing this dynamism of an executing system incorporating new reconfiguration processes that were not initially incorporated at design time. New reconfiguration or management processes are added to the platform an acceptance test is passed.

A generic model is presented in this work based on [23] that can be applied in any dynamic framework, based on real-time or for QoS adaptation purposes. Enough resources are reserved to keep the QoS properties and feasibility of the system during the management process. This process is executed under temporal control so that a maximum time bound is achieved.

## 3    System model and infrastructure

This section describes the component-based system model and the infrastructure developed for supporting run-time modifications of supported operations and of individual components.

### 3.1    Component-based model

We consider that the functionality in the system is provided as components. All components are considered to be active and periodic. This means that each component contains an execution thread that is activated periodically. Therefore, temporal properties of components are:

- $C_i$ that is the *worst case execution time* (WCET) of the thread of component *i*.
- $T_i$ or *activation period* of component *i*, that is the time between each two consecutive activations.
- $D_i$ or relative *deadline* for component *i*.

Since components have periodic activations, the cyclic task model [10] can be applied. In this way, *rate monotonic analysis*(RMS) [9] is used in the temporal scheduling of the component substitutions. The priority is assigned according to the activation frequency of the component in such a way that components with higher activation frequency also receive highest priority. It would also be possible to consider *earliest deadline first* (EDF) [10], but the schedulability model for EDF would add extra complexity if the necessary restrictions are not kept [4]. The execution of the operations required for the substitution should not be interrupted. However, EDF modifies priorities dynamically, and it is used with pre-emption mechanisms. In the absence of preemption, EDF would lack one of its main claims over RMS, which is theoretically reaching a 100% CPU occupancy. Dynamic priorities and preemption may result in the interruption of the replacement operations because of the priority some task or component being raised.

The number of running components in the system is considered to be dynamic. Periodically, new components can be loaded to memory, and they can start execution supporting a controlled level of *free execution*. Therefore, in this work, free execution refers to the capacity of supporting run-time dynamism via new execution, substitution, or removal of components.

### 3.2 Infrastructure and management operations

Run-time substitution (including addition or removal) of a component requires the execution of a set of *management operations* with the goal of performing the required updates to the structure of connections and execution status of components. Such operations are provided inside an infrastructure as shown in **Fig. 1**.

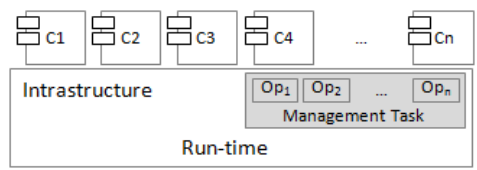

**Fig. 1.** Overview of the infrastructure

Different execution infrastructures can be offered for this purpose containing a number of *management tasks* in charge of performing the required management operations. The presented infrastructure has the objective of being simple and efficient. Therefore, it contains a single management task that takes control every time a component has to be replaced. The management operations it performs in every replacement are the following:

- *Adding a new component*. The component is loaded to main memory and connected to other components.
- *Removing a component*. A running component is stopped and *disconnected* from the system. It is unloaded from main memory.
- *Modification a component*. The execution parameters of a component are modified to change its behavior.

- *Modification of connections*. The connections or *bindings* between components are modified to change the behavior of the application by updating the execution and invocation flows.
- *Replacement of a component*. An active component is replaced by a new instance to modify its behavior. This is, for example, needed when an erroneous component needs to be replaced to fix a bug.

Requests for management operations follow a *sporadic* pattern. The user or any running component in the system may issue a request for one of these management operations to be performed. Requests are queued waiting for the management task to execute and to perform the corresponding management operation in the system (see **Fig. 2**).

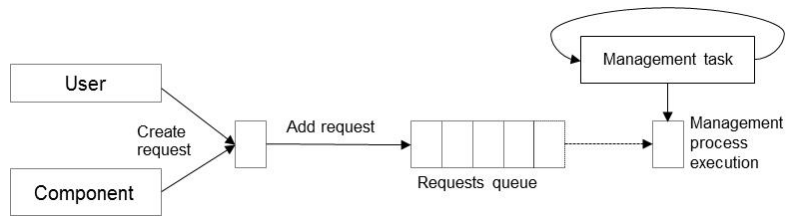

**Fig. 2.** Management task execution

The number of management operations can differ in different infrastructures depending on the requirements of the specific application domain (see **Fig. 3**).

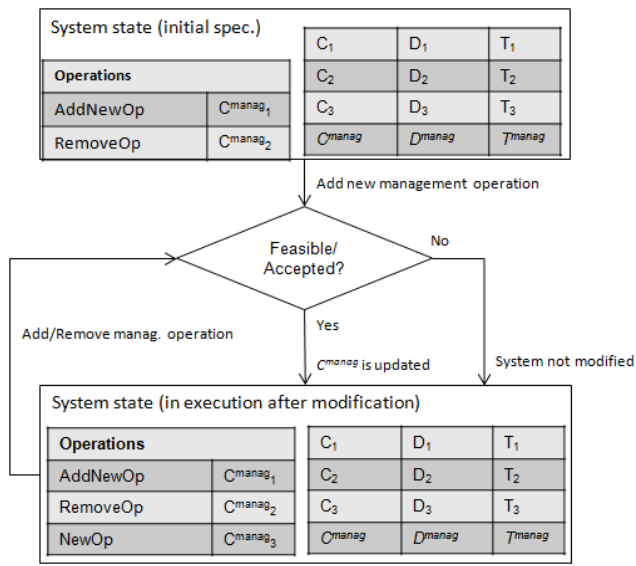

**Fig. 3.** Run-time variation of management operations

For instance, in a hard real-time context, it may be the case that only the removal of a component is allowed; whereas in a soft real-time domain, all of the above and possibly some additional operations may be allowed.

### 3.3 Execution model

The execution of management operations has to be scheduled to avoid race conditions and interference during their execution that could result in incoherent internal states of components. The following premises must hold to guarantee the absence of interference in the execution of the management operations:

- Enough processor time has been reserved for the execution of the management operations. A schedulability analysis can be, then, applied to ensure temporal isolation in their execution.
- Components affected by the management process must not run concurrently with the management process. We assume that components to be replaced stop at safe execution points, i.e., in a safe point, the state of a component is stable.

Although management processes are supposed to be requested sporadically, the use of a periodic task is proposed in this infrastructure for the correct schedulability of the management operations. The management task $\tau^{manag}$ is in charge of applying a management processes when possible, i.e., at the next activation period. The timing parameters of the task are:

- *Activation period* ($T^{manag}$). The execution period of the task has to be fine-tuned according to the application needs. This value matches the minimum inter-arrival time (MIT) of management operations expected during the execution of the application.
- *Computation time* ($C^{manag}$). This value corresponds to the WCET of the available management processes. Considering a number of management processes that can be applied in the system as $C^{manag}_1$, $C^{manag}_2$, $C^{manag}_3$, etc., then, the larger value is used.

$$C^{manag} = \max(C^{manag}_1, ..., C^{manag}_n) \qquad (1)$$

being $n$ the number of management operations available in the system, and $C^{manag}_j$ is the cost of execution of the management operation $j$ in the system.

- *Deadline* ($D^{manag}$). The deadline to finish the management operation. It is set to the same value as the cost ($C^{manag}$) to enforce the completion of the operation before the execution of any other component continues.

**Fig. 4** shows the execution of two component tasks ($co_1$ and $co_2$) and the management task, $\tau^{manag}$, and the replacement of component $co_2$. Since the management task has lowest priority, $\tau^{manag}$ is interrupted by any other higher priority task. In this case, it can be seen that $co_2$ interrupts $\tau^{manag}$. When the component $co_2$ resumes its execution, it may be in a corrupted state as the modification of the component may not be completed. This situation results in anomalous executions that may leave the system in an unstable state.

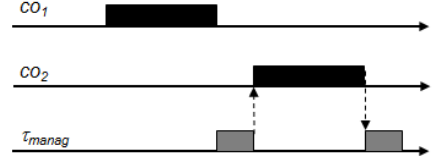

**Fig. 4.** Interfered *execution of a management operation*

In our framework, management operations can only take place at predefined time instants when the management task executes. Therefore, a limited number of management operations can be applied during a given time interval. This way, resources assigned to the management task may be considerably reduced, which has a minimum impact on the schedulability of the rest of system tasks.

## 4      Admission control for management operations

Integrating the proposed infrastructure that contains a *management task* has an impact on the schedulability of the system as a whole. Following, the exact response time analysis of Lehoczky et al. [9] is applied to evaluate this implication.

$$R_i = C_i + \sum_{j=hp(i)} \left\lceil \frac{R_i}{T_j} \right\rceil C_j \qquad (2)$$

Equation (2) reflects the cost of components. The infrastructure cost should be considered also as part of the interference since the management task will have higher priority than any other component belonging to the set of *hp(i)*. In the proposed model, the safe execution of component substitutions is achieved in the following way. The effect of the execution of $\tau^{manag}$ with the highest priority in the schedulability analysis of the task set is obtained from a modification of Equation (2) yielding Equation (3):

$$R_i = C_i + \sum_{j=hp(i)} \left\lceil \frac{R_i}{T_j} \right\rceil C_j + \left\lceil \frac{R_i}{T^{manag}} \right\rceil C^{manag} \qquad (3)$$

where the interference created by the execution of the management task is added to the response time of every task in the system (see **Fig. 5**). The properties of this task will be updated according to the modifications of the system, when new management processes are added or removed.

The processor time reserved for $\tau^{manag}$, i.e., the value of $C^{manag}$, is application dependant. The baseline to determine such a value is Equation (2) to assess the impact of the addition of the management task in the schedulability of the whole system. In this context, the proposed infrastructure assigns the highest priority in the system to

$\tau^{manag}$ so that it cannot be interrupted during the execution of a management operation. Parameters of $\tau^{manag}$ are equivalent to other component tasks as computation time ($C^{manag}$), deadline ($D^{manag}$), and its activation period ($T^{manag}$). These are fine tuned off-line according to the application needs.

**Fig. 5** shows the execution model of the management task $\tau^{manag}$. This execution model is given by Equation (3). Only one management process can be applied when this task is scheduled. The number of management processes that can be performed in the system is tuned by modifying the activation period of task $\tau^{manag}$.

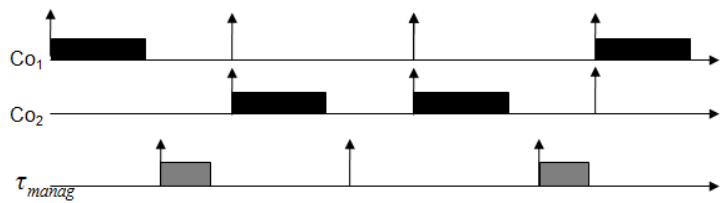

**Fig. 5.** Periodic execution of the management task for component substitution

The period of the new task $\tau^{manag}$ will be determined by the *processor reservation time* or required *minimum inter-arrival time (MIT)* for management operations. Two ways to calculate $T^{manag}$ are proposed here.

To reserve time for aperiodic or background, the percentage of processor time can be calculated to obtain the approximated execution period of the management task. Then, this task's period is given by:

$$T^{manag} = \frac{100 * C_m}{U} \quad (4)$$

Where $U$ is the percentage of CPU utilization reserved for the task.

A minimum number of management operations can be expected in the platform at run-time during a specified time. This number will be given by the allowed MIT requests for component execution changes. This minimum inter-arrival time will directly determine the period of this task. If we want to reserve enough processor time for a number of management processes, $n_{manag}$, in a fixed period of time, $t$, then the minimum inter-arrival time, and hence the period of the management task, will be given by:

$$T^{manag} = \frac{t}{n_{manag}} \quad (5)$$

Additionally, other techniques can be used to reduce the interference of this task, i.e., using a periodic execution value equal to one of an already existing task in the system. A tolerance value is used for determining how close the values are. If the task with the closest period is used, then the schedulability analysis is simplified compared to the arbitrary selection of periods.

To incorporate new management operations at run-time an on-line acceptance test is required that will be based on a schedulability analysis using the equations described in the previous section. Adding new management operations may increase the computational cost of the management task since the number of instructions to be executed by it will increase. Processor time required by the management task is calculated using Equation (1). For the schedulability analysis, Equations (3) is applied. If the system is still schedulable after introducing those changes, then the new management operation is accepted. **Fig. 6** shows the acceptance test.

```
C^manag=max(C^manag_1, …,C^manag_n)
rc=calculate_RTA_schedulability(TS);
if(rc==true)then
        accept_operation();
elsebegin
        reject_operation();
end;
```

**Fig. 6.** Acceptance test for replacement

*TS* refers to the current task set in the system that includes $\tau^{manag}$. The value of $C^{manag}$ for $\tau^{manag}$ in the task set is calculated and modified if needed. The schedulability analysis is applied. If the test is passed, the management operation is accepted; else, the operation is rejected, and the task set keeps its previous state.

As an improvement, this operation can be avoided if its computation time of the operation that is checked is smaller than or equal to the computation time of the management task ($C_j^{manag} \leq C^{manag}$), as there is enough time reserved for its execution.

It should be noted that additional requirements may be necessary for the application of some management operations, and the corresponding check can be added in this acceptance test. Additional requirements are considered out of the scope of this work.

## 5    Model implementation and validation

A high level view of the infrastructure has been elaborated as a structural diagram in UML that offers a direct path to implementation (see **Fig. 7**).

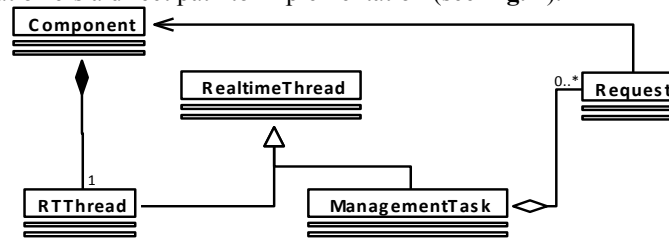

**Fig. 7.** Realization of the infrastructure. Structural view.

The first element to implement is the component specification (template or class) representing the basic functional units. Each component functionality is executed by a real-time thread. Every `Component` specification contains `RTThread` unit for such purpose. The `RTThread` extends and simplifies a `RealtimeThread` class that provides all the real-time characteristics of a real-time periodic thread. `RTThread` offers a simplification to easily handle the periodic component functionality. Temporal properties as periods, WCET, and deadlines are provided to the class constructor for it to be correctly scheduled. An abstract `execute` function is implemented in every component to provide the functionality. This method will be executed according to the temporal properties previously provided. A `wrapup` function is provided for clean-up when a component is stopped by freeing resources and connections if necessary.

The `ManagementTask` specification is provided that inherits the `RealtimeThread` specification. It is implemented with the previously defined real-time parameters. It also has access to all the available management operation requests.

Operation requests are queued until the management task. A hierarchy of request can be implemented based on a generic `Request` specification. **Fig. 8** shows an example of hierarchy of requests.

Requests may be implemented for every desired management process, as those described in previous section, i.e., loading a component, reconfiguring a component, reconfiguring connections, etc. The `apply` method is expected to implement the corresponding management process.

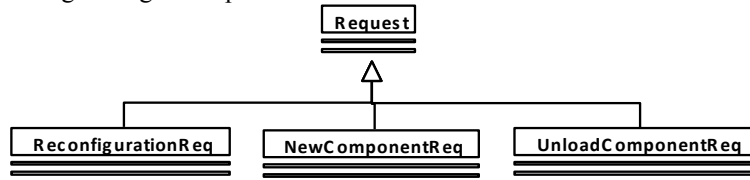

**Fig. 8.** Hierarchy of Requests

Two management processes could be considered essential: the loading and unloading of management processes. The load of new code into main memory is considered out of the scope of this work.

*Prototype implementation*

Our method is intended to be applied to more dynamic component systems, specifically the one provided by the OSGi platform [13], taking into account the limitations and requirements of RTSJ [2] such as the garbage collector [18].

The presented infrastructure offers a set of capabilities that are not completely fulfilled by the related work (see **Table 1**). OSGi offers a dynamism that is not offered by any other platform; components can be added, removed, replaced, and reconfigured on-line. But no support for real-time capabilities is provided. The management operations are not bounded in time and not suitable for real-time applications.

The Rasche and Polze [15, 16] proposals provide a process to safely apply reconfigurations on the application on-line. These reconfigurations may imply the replacement of components. The capability to add new reconfiguration operations while the system is in execution is not detailed in their work. The described process, based on halting the execution of running components, do not seem suitable for real-time applications as the execution deadlines of tasks would result affected.

Qoskets [20] are designed to apply on-line reconfiguration of components and with some support real-time characteristics. The reconfiguration capabilities are encapsulated with the components, and no support for replacements is described. This limits the possible reconfigurations to those initially in the platform. There is no support for the addition of new management operations at run-time.

The infrastructure described here is designed to allow the incorporation of new management operations while the system is running. The characteristics of the added management operation are not previously known by the infrastructure.

The safety is provided by the enforcement of time bounds in the execution of management operations by the management task. New management operations are accepted only if the system is schedulable.

| Technology | Support for modification of components | Real-time Support | Addition of new operations |
|---|---|---|---|
| OSGi | Yes (bundles) | No | Yes |
| Rasche & Polze | Yes (components) | No | No |
| Qoskets (CORBA) | No | Yes | No |
| Infrastructure | Yes (components) | Yes | Yes |

**Table 1.** Comparison of infrastructure characteristics

Stress tests where applied in a Java implementation to check the viability of the insfrastructure applying management operations using Jamaica[1] virtual machine. The test is applied generating a set of 100 running components. A single management operation is used by the infrastructure in charge of substituting components. This operation is continuously being requested to randomly substitute running components during 120 seconds. The resulting average time required by the substitution to complete is 0.242*ms* while the maximum time required for a component replacement is 1.559*ms*.

This shows the interference generated by the use of a virtual machine and the garbage collector of a Java environment.

## 6   Conclusion

In this work, it is described a simple component-based model with the associated execution infrastructure to achieve the safe execution of management operations that increase the degree of freedom in the execution of the system. The infrastructure integrates the required temporal properties to enable schedulability analysis for assessing the possibility of executing new management operations. Processor time is reserved for the management operations in the overall processor reservation calculations. To avoid erroneous execution of management operations (e.g. due to their execution by a background priority task), an acceptance test is proposed that enables the usage of cost estimations for the management operations. Also, a periodic task that is assigned a reserved time slot equivalent to the largest management operation in the system. This time slot is calculated pessimistically to ensure safe execution, but it can be used to perform any of the management operations. The periodicity of the execution of this time slot will depend on the management operations needed by the application.

---

[1] http://www.aicas.com/jamaica.html, December, 2012.

We validate the idea by providing a model for its implementation, and also by providing results of an actual implementation for dynamic soft real-time systems based on OSGi showing that components can be safely modified using this method.

**Acknowledgments.** This work has been partly supported by the iLAND project (ARTEMIS-JU 100026) funded by the ARTEMIS JTU Call 1 and the Spanish Ministry of Industry, Commerce, and Tourism (www.iland-artemis.org), ARTISTDesign NoE (IST-2007-214373) of the EU 7th Framework Programme, and by the Spanish national project REM4VSS (TIN 2011-28339).